\documentclass[aps]{revtex4}

\usepackage{amssymb}
\usepackage{amsmath}
\usepackage{graphicx}
\usepackage{float}

\begin{document}

%*********************************************************************
\title{Optimal smoothing parameter in Eilers-Whittaker smoother}
%**********************************************************************

\author{R. \surname{Bernal-Arencibia}}
\affiliation{MATCOM, Universidad de La Habana, San Lazaro y L. CP 10400. La Habana. Cuba.}

\author{K. \surname{Garcia-Medina}}
\affiliation{Facultad de F\'isica, Universidad de La Habana, San Lazaro y L. CP 10400. La Habana. Cuba.}

\author{E. \surname{Estevez-Rams}}
\email{estevez@fisica.uh.cu}
\affiliation{Facultad de F\'isica-Instituto de Ciencias y Tecnolog\'ia de Materiales(IMRE), Universidad de La Habana, San Lazaro y L. CP 10400. La Habana. Cuba.}

\author{B. \surname{Aragon-Fernandez}}
\affiliation{Parque Cientifico Tecnologico de La Habana. Carretera a San Antonio km 2 1/2. Boyeros. La Habana.}

\begin{abstract}
The Eilers-Whittaker smoother's effectiveness depends on the choice of the regularization parameter, $\lambda$, and automatic selection is a necessity for large datasets. Common methods like leave-one-out cross-validation can perform poorly with serially correlated noise. We propose a novel procedure for selecting $\lambda$ based on the spectral entropy of the signal's residuals.

We define an S-curve from the Euclidean distance between points in a plot of the spectral entropy of the residuals versus that of the smoothed signal. The $\lambda$ corresponding to the absolute maximum of this S-curve is chosen as the optimal parameter. Using simulated data, we benchmarked our method against cross-validation and the V-curve. Validation on diverse experimental data was also done. This simple and robust procedure can be a valuable addition to the available selection methods for the Eilers smoother.
\end{abstract}

%\begin{keyword}
 %noise \sep smoothing \sep Eilers-Wittaker algorithm
%\end{keyword}

\date{\today}
\maketitle                        % DO NOT DELETE THIS LINE

%................................................................................................
\section{Introduction}

Data processing is a cornerstone of experimental work. Science is an endeavour driven by experimental facts, and those are delivered in the form of data. However, as everyone involved in experimental work knows, data rarely, if ever, comes free from noise corruption. Noise, in the sense used here, refers to all non-systematic factors that deviate in a statistically random manner from the signal-recording process, resulting in data that only approximates the underlying regularities or laws governing the studied phenomenon. In fields such as engineering, physics, and the life sciences, noise can obscure meaningful patterns, driving the need for smoothing techniques that enhance the signal-to-noise ratio without compromising data integrity. Over time, various statistical approaches have been developed to address this challenge, including classical moving-average filters \cite{smith97}, exponential smoothing \cite{hyndman02}, regularisation-based methods, nonlinear methods \cite{schreiber91,kantz93}, and, more recently, deep learning methods \cite{oppliger24}.

Eilers introduced a smoothing procedure based on a linear optimization procedure with regularisation \cite{eilers03}. In the so-called Eilers-Whittaker method, two terms are used in a minimization procedure, one takes care of reducing the mean square error between the smoothed data and the original signal. In contrast, the second regularisation term is a penalty term that intends to smooth the solution, effectively reducing noise at the expense of the first term. A regularisation parameter governs the balance between the two terms. The selection of the optimal regularisation parameter remains a critical issue: excessive values may suppress essential features, while insufficient values fail to effectively attenuate noise.

At the heart of the regularisation parameter selection is the ambiguous statement of the noise reduction problem. By definition, there is no law governing noise $n(t)$, while the uncorrupted signal itself $s(t)$ is not known beforehand. An often used model for describing the experimental result is
\begin{equation}
 y(t)=s(t)+\lambda n(t), \label{eq:signal}
\end{equation}
where $y(t)$ is the experimental data and $\lambda$ governs the amplitude of the noise. Although the statistical model of noise is generally unknown, when simulating corrupted data, it is assumed to follow some prescribed distribution. The most commonly used is the so-called white noise with a Gaussian distribution, or, in other cases, the less realistic uniform noise. In any case, noise reduction is how to get from a signal $y(t)$ to data as close as possible to $s(t)$ with the least assumption possible regarding the nature of $n(t)$.

The most general assumption that can be made of noise is that, compared to the signal, it has a less smooth behaviour, or in Fourier terms, it has a more prominent higher frequency component than the signal $s(t)$. One should expect that, when subtracting the experimental data from the smoothed data, the remaining 'noise' should be unbiased, optimally with a near-zero mean and a random distribution around the mean value. However, these two conditions can be achieved by a large range of parameters governing the smoothing procedure. In view of this fact, when confronted with the smoothing problem, a common piece of advice is to 'trust' your instinct and, upon visualizing the experimental data and the smoothed signal, subjectively choose the parameters that seem to fit the signal best. Such a criterion, while feasible, despite its shortcomings, becomes unrealistic when dealing with large sets of data, where automatic selection without supervision becomes a necessity.

In the original work of Eilers, parameter selection was carried out through leave-one-out cross-validation \cite{eilers03}. In this procedure, a point of the experimental data is left out, the remaining points are smoothed, and a prediction is made for the missing one. By repeating the procedure for each experimental point, a final standard error can be computed between the predicted data and the experimental data. The regularization parameter that minimizes the standard cross-validation error is chosen as the optimal one. There are efficient methods for computing the cross-validation error, which significantly reduce the computational effort. In the not uncommon case of correlation in the noise, leave-one-out cross-validation tends to choose a parameter that underfits the data. The reason for this is that a smooth signal plus correlated noise can be seen as a less smooth signal plus white noise. Similar problems can be found in other selection procedures based on generalized cross-validation \cite{golub79,golub97}, Akaike's information criterion \cite{zhang10}, or Bayesian information criterion \cite{huri16}.

To overcome the limitations of these methods, Frasso and Eilers \cite{frasso15} proposed using the L- and V-curves, which were first introduced by Hansen et al. \cite{hansen93}. The L-curve is a plot of the logarithm of the magnitude of the penalty term in the Eilers minimization function, against the logarithm of the sum of the squared residuals, parametrized by the regularization parameter. The regularization term is chosen where a corner in the L-curve appears. The method proves to be less influenced by correlation in the noise in several cases. As the location of the corner can be a costly computation, a transformation of the L-curve is therefore done to find, alternatively, the two points that are closest together in the L-curve by measuring their geometric mean. The curve of such distance against the parametrization parameter is called the V-curve due to its shape; the minimum of the curve determines the optimal regularization parameter.

In this contribution, we explore a new parameter estimation procedure based on the spectral entropy of the signal. The method is simple yet exhibits robust and efficient behaviour in a wide variety of simulated and experimental cases. The reported procedure can be added to the toolkit of selection procedures available for the Eilers smoothing method.

%................................................................................................
\section{Eilers-Whittaker algorithm}

The starting point is noisy data $y(t)$, which is the experimental realization of an underlying smooth signal $s(t)$ corrupted with noise as described by equation (\ref{eq:signal}). The residual of the experimental data and the underlying smooth data will be given by
\begin{equation}
 R=\sum\limits_t^n (y(t)-\hat{s}(t))^2,\label{eq:R}
\end{equation}
whereas $\hat{s}(t)$ is an approximation of $s(t)$ upon smoothing. $n$ is the number of experimental sampled points. Smoothing can be measured by the n-order numerical derivative of the experimental data. Up to the first order, this will be given by
\begin{equation}
 S_1=\sum\limits_q^n (\hat{s}(t)-\hat{s}(t-1))^2,\label{eq:S}
\end{equation}
while for second-order derivatives the corresponding expression would be
\begin{equation}
 S_2=\sum\limits_q^n (\hat{s}(t)-2\hat{s}(t-1)+\hat{s}(t-2))^2,
\end{equation}
and so on.

A compromise is taken between achieving a small value for $R$ and for $S$. The smoothing procedure is set as an optimization problem that minimizes the metric
\begin{equation}
 Q=R+\lambda S_o \label{eq:Q}
\end{equation}
where $\lambda$ is the regularization parameter that decides the balance between $R$ and $S_o$ minimization. The bigger $\lambda$, the more the minimization will be driven by the $S$ factor and larger smoothing of the signal will happen; on the other side, the smaller the $\lambda$ value, the closer to the noisy signal $y(t)$ the solution $\hat{s}$ will be.

Writing $|x\rangle=(x(q_1),x(q_2)\ldots, x(q_n) )$ and calling $\langle x| x\rangle=\sum_ q |x(q)|^2$, the internal product of vector $|x\rangle$, we can write equation (\ref{eq:Q}) as
\begin{equation}
\begin{array}{ll}
 Q&=\langle y-\hat{s}|y-\hat{s}\rangle+\lambda \langle \Delta \hat{s}|\Delta \hat{s}\rangle\\\\
 &=\langle y-\hat{s}|y-\hat{s}\rangle+\lambda\langle \hat{s}| D^T D|\hat{s}\rangle,
 \end{array}
\end{equation}
where, for first order differences, $D$ is a $(n-1)\times n$ matrix,
\begin{equation}
 D=\left (\begin{array}{ccccccc}
 -1 & 1 & 0 & 0 & \ldots & 0 & 0 \\
 0 & -1 & 1 & 0 & \ldots & 0 & 0 \\
 0 & 0 & -1 & 1 & \ldots & 0 & 0 \\
   &   &  \cdot  & \cdot  & \cdot &   &   \\
 0 & 0 & 0  & 0 & \ldots & -1 & 1
   \end{array}\right),
\end{equation}
and for second order, the $(n-2)\times n$ matrix,
\begin{equation}
 D=\left (\begin{array}{ccccccccc}
 1 & -2 & 1 & 0 & 0 & \ldots & 0  & 0 & 0\\
 0 & 1 & -2 & 1 & 0 & \ldots & 0 & 0 & 0\\
 0 & 0 & 0 & 1 & -2 & \ldots & 0  & 0 & 0\\
   & &   &  \cdot  & \cdot  & \cdot &   &   \\
 0 & 0 & 0 & 0 & 0 & \ldots & 1& -2 & 1
   \end{array}\right),
\end{equation}
The $|\hat{s}\rangle$ vector which minimizes $Q$ will be given by $\partial Q/\partial |\hat{s}\rangle=0$,
\begin{equation}
 -|y-\hat{s}\rangle+\lambda D^T D |\hat{s}\rangle=0
\end{equation}
which can be written as a linear system of equations
\begin{equation}
 W |\hat{s}\rangle=|y\rangle,\label{eq:w}
\end{equation}
where, taking $I$ as the identity matrix,
\begin{equation}
 W=I+\lambda D^T D.
\end{equation}
It is useful to introduce the hat matrix
\begin{equation}
H= (I + \lambda D^T D)^{-1}.\label{eq:hat}
\end{equation}

The solution to the system (\ref{eq:w}) will be the smooth function (controlled by the parameter $\lambda$) derived from $y(t)$.

%................................................................................................
\section{The regularization parameter}

The $\lambda$ parameter in expression (\ref{eq:Q}) determines the smoothing behaviour of the Eilers algorithm and therefore, how to choose it becomes an important step in the smoothing procedure. The goal of selection procedures is to provide a systematic method for choosing the smoothing parameter, $\lambda$, rather than relying on subjective visual assessment. Several procedures are used for this selection, and, for completeness, we describe two of the most commonly used ones.

\subsection{Leave-one-out cross-validation}

The fundamental idea behind this procedure is to iteratively exclude each data point, smooth the remaining data, and predict the value of the omitted point. This process is repeated for all non-missing data points. The cross-validation standard error, $\sigma_{cv}$, is then computed using the following formula:
\[ \sigma_{cv} = \sqrt{\frac{\sum_{i} (y_i - \hat{y}_{-i})^2}{m}}, \]
where $y_i$ is the observed data point, $\hat{y}_{-i}$ is its prediction when it was left out of the smoothing process, and $m$ is the number of non-missing observations. To find the optimal $\lambda$, this entire procedure is performed over a grid of $\lambda$ values (typically spaced logarithmically), and the value that minimizes $\sigma_{cv}$ is chosen.

A literal implementation of the procedure is computationally costly, with a complexity proportional to $m^2$. A much faster approach, proportional to $m$, is possible by using a well-known property of the hat matrix, $H$. The key relationship allows the calculation of cross-validation residuals directly from standard residuals and the diagonal elements of the hat matrix, $h_{ii}$:
\[
y_i - \hat{y}_{-i} = \frac{y_i - \hat{y}_i}{1 - h_{ii}}.
\]
This formula makes the calculation trivial if the diagonal of $H$ is known. However, computing the full matrix $H$ is impractical for large datasets. To solve the computation of the $H$ matrix, it must be observed that the shape of the diagonal elements $h_{ii}$ is nearly identical for problems of different sizes, provided that $\lambda$ is scaled appropriately. This allows the diagonal of $H$ for a very large dataset to be accurately estimated by computing the full $H$ for a much smaller problem, this fact combined with fast diagonal estimation like the Hutchinson methos \cite{dha23}, leads to fast cross-validation \cite{eilers03}.

Standard cross-validation assumes that the errors in the data are independent. If the errors exhibit serial correlation, cross-validation tends to select a value of $\lambda$ that is too small, resulting in an under-smoothed curve.

\subsection{L- and V-curve}

The L-curve is a tool initially developed for ill-posed inverse problems that can be effectively applied to smoothing. It provides a graphical way to analyze the trade-off between the two core components of a smoothing procedure: the badness of the fit (residuals) and the roughness of the estimate. The L-curve is a parametric plot of the logarithm of the penalty term (first term $R$ in equation (\ref{eq:Q})) versus the logarithm of the sum of squared residuals (second term $S$ in equation (\ref{eq:Q})). Plotting the points $(\log R, \log S)$ yields a curve typically with a distinct 'L' shape.

The optimal smoothing parameter $\lambda$ is found at the ''corner'' of the L-curve, which represents the point of maximum curvature. This corner is were a balanced compromise is achieved, where the relative changes in the size of the residuals are approximately equal to the relative changes in the penalty term. The procedure is insensitive to serial correlation in the noise, a common issue that can cause cross-validation to fail.

As the computation of the ''corner'' can be computationally cumbersome, the V-curve is introduced, based on the observation that the points plotted on the L-curve are most closely spaced near the corner. The V-curve method calculates the Euclidean distance between consecutive points in the L-curve. This distance is then plotted against the geometric mean of their corresponding $\lambda$ values. The resulting plot typically has a ''V'' shape. The optimal smoothing parameter $\lambda$ is chosen as the value that corresponds to the minimum of the curve. This minimum is located very close to the point of maximum curvature on the L-curve. The V-curve generally chooses the same optimal $\lambda$ values as the L-curve and performs well in simulations and real-data applications.

%................................................................................................
\section{Spectral entropy selection criterion}

\begin{figure}[!t]
\centering
\includegraphics[scale=0.5]{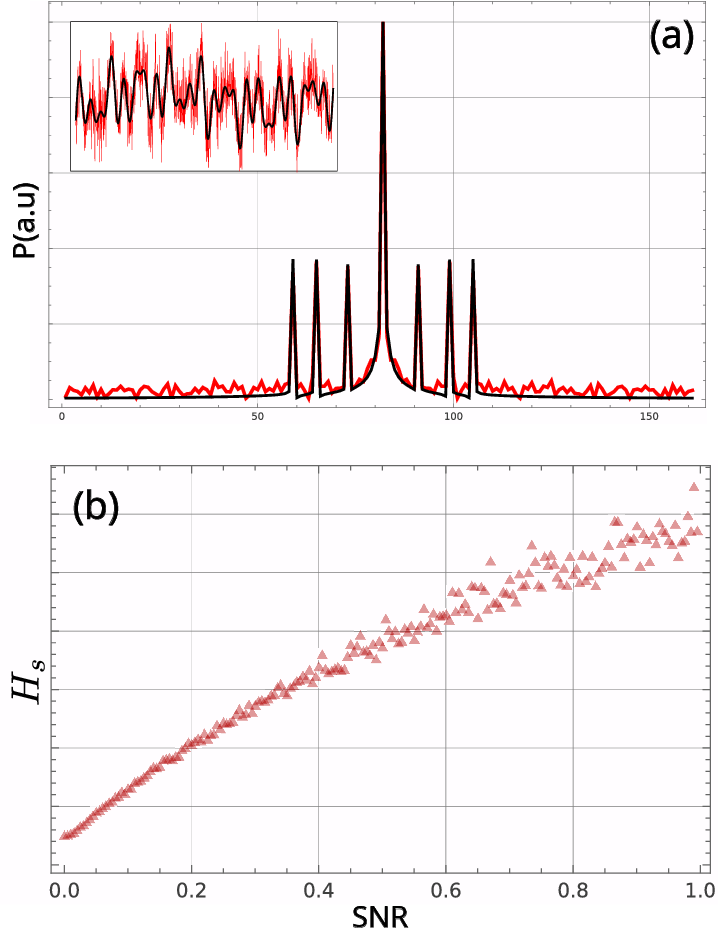}
\caption{\textbf{Spectral entropy.} (a) Power spectrum $P(q)$ of an arbitrary function ($f(t)=1/4(\sin t+\sin 9 t+\sin 17 t+\sin 23 t+\log(t+1))$) without noise (black) and with Gaussian noise (red). The inset shows the function plot. Noise adds a continuous spectrum to the Fourier transform, while the signal without noise shows a compact support in the middle range. (b) The Shannon entropy $H_s$ over the normalized power spectrum as a function of noise level (variance in the Gaussian noise distribution). $H_s$ monotonically increases with the signal-to-noise ratio. }\label{fig:noisesignal}
\end{figure}

Take a signal $f(t)$ and consider its Fourier transform $\mathcal{F}(q)$. For a signal dominated by random noise, the Fourier spectrum should be dense and cover the whole range of $q$ values, for a smooth function one should expect that the support of the power spectrum will be finite and, ideally, if the signal is smooth enough, the higher frequencies should be zero (Figure \ref{fig:noisesignal}). The corresponding power spectrum of the Fourier transform after normalization by its sum can be used to calculate an entropy \cite{powell79,stan09} like magnitude through
\begin{equation}
 H_S=-\sum\limits_{q=-\infty}^\infty F(q)\log F(q),
\end{equation}
where $F(q)=|\mathcal{F}(q)|^2$ is the normalized magnitude of the complex component of the Fourier transform for frequency $q$, such that $\sum F(q)=1$. $H_S$ is known as the spectral entropy.

For each $\lambda_i$ value, take the residuals given by equation (\ref{eq:R}) and calculate their spectral entropy. Perform the same calculation for $D\hat{s}$, that is, the difference operator acting over the smoothed data. Let us call such a value $H_{\hat{y}}$. The smoother the solution, the smaller the high-frequency bound of the power spectrum. We proceed as follows. Inspired by the L- and V- curve construction, consider the set of $h_{\lambda_i}=(\log H_S, \log H_{\hat{y}})$, and from there calculated a V-type curve by taking consecutive values and computing its Euclidean distance

\begin{equation}
e_{\lambda_i}=\| h_{\lambda_i+1}-h_{\lambda_i}\|
\end{equation}

\begin{figure}[!t]
\centering
\includegraphics[scale=0.5]{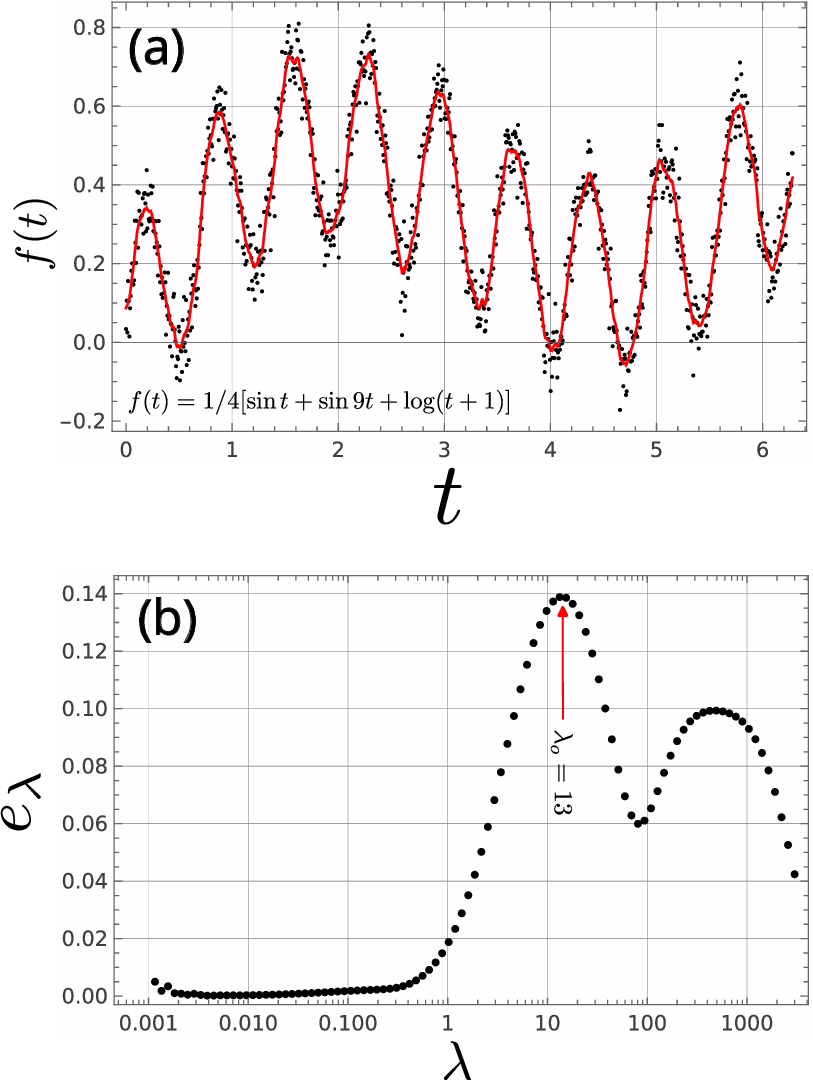}
\caption{\textbf{Spectral selection.} (a) Arbitrary analytical function with noise. (b) The S-curve for the spectral analysis of the residuals (see text for explanation). The optimal $\lambda$ value ($\lambda_o$) is chosen as the value where the S-curve has an absolute maximum. }\label{fig:vcurve}
\end{figure}

The corresponding $e_\lambda$ vs $\lambda$ curve will be called the S-curve. In Figure \ref{fig:vcurve}, the S-curve of an arbitrary continuous function is shown. The curve shows local maximum values. The $\lambda$ value for the absolute maximum of $e_\lambda$ was empirically found to be a good choice for the regularization parameter.

%................................................................................................
\section{Results and Discussion}

To compare the spectral entropy selection criterion with the cross-validation and the V-curve procedure, we take an analytical function $s(t)$, and add some Gaussian noise of a given level to get a noisy function $y(t)$. For a given $\lambda$ value the smoothed function $\hat{s}(t)$ is obtained, the mean square error ($mse$) between both functions $\sum |s(t)-\hat{s}(t)|$ is computed. The procedure is repeated, covering an interval of $\lambda$ values. The $\lambda_o$ value with the smallest mean-square-error $(mse)_o$ is chosen as the optimal regularization parameter. This value is compared with the $\lambda$ value chosen by each selection procedure, namely, cross-validation, V-curve and S-curve selection. The procedure is repeated for different noise levels.

\begin{figure}[!t]
\centering
\includegraphics[scale=0.8]{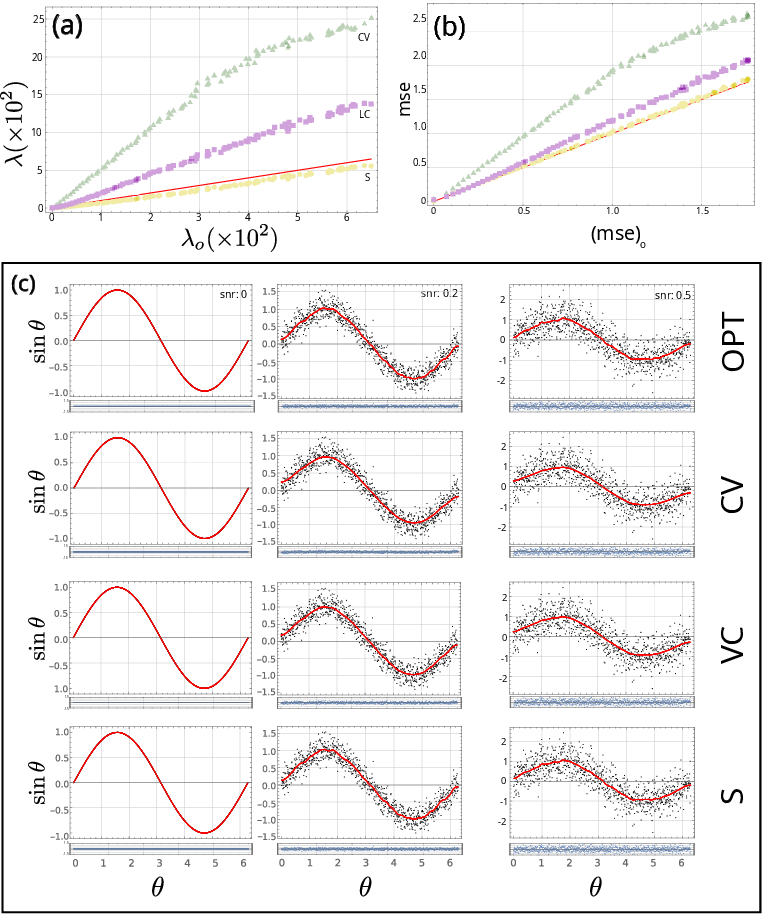}
\caption{\textbf{Noisy sinus function.} (a) The $\lambda$ value chosen by each method versus the optimal $\lambda$ ($\lambda_o$). The optimal regularization parameter is the one that produces the minimum mean square error $(mse)$ between the smoothed signal and the original, uncorrupted signal. The red line represents the ideal case where the selected $\lambda$ equals $\lambda_o$. (b) Comparison between the mean square error $(mse$) produced by each method's chosen $\lambda$ against the optimal mse ($(mse)_o$). The red line indicates optimal performance. (c) A visual comparison of the smoothed curves generated by each method (CV, VC, and S) and the optimal smoother (OPT) for three different signal-to-noise ratios (snr: 0, 0.2, and 0.5). While visual inspection makes it difficult to distinguish significant differences between the methods, the quantitative analysis in the upper panels confirms the superior performance of the Spectral Entropy approach.}
\label{fig:sinus}
\end{figure}

Figure \ref{fig:sinus} shows the result for a $\sin (t)$ function. In the lower section of the figure, the smoothed curve using the different selection criteria is compared to the optimal curve (OPT). The visual inspection of the smoothed curves with the different selection criteria proves difficult to distinguish among them, which points to the need for a quantitative analysis. The upper left plot compares the $\lambda$ for each selection method with the optimal regularization parameter. The solid red line is the optimal $\lambda_o$ value. The regularization parameter increases with the noise level; therefore, in the plot, increasing $\lambda_o$ values corresponds to a decreasing signal-to-noise ratio. As can be seen, the cross-validation and V-curve criteria tend to overestimate the value of the regularization parameter; this behaviour worsens with increasing noise level (increasing $\lambda_o$). Cross-validation is the worst-performing method, while the V-curve is better; however, both are significantly worse than spectral entropy selection, which nearly follows the optimal $\lambda$ line. The result can also be observed when comparing the mean-square error, as shown in the upper right plot of the same figure. The worst $mse$ behaviour is that of cross-validation, while the spectral entropy procedure gives a $mse$ close to the optimal one in the whole range of noise values.

\begin{figure}[!t]
\centering
\includegraphics[scale=0.8]{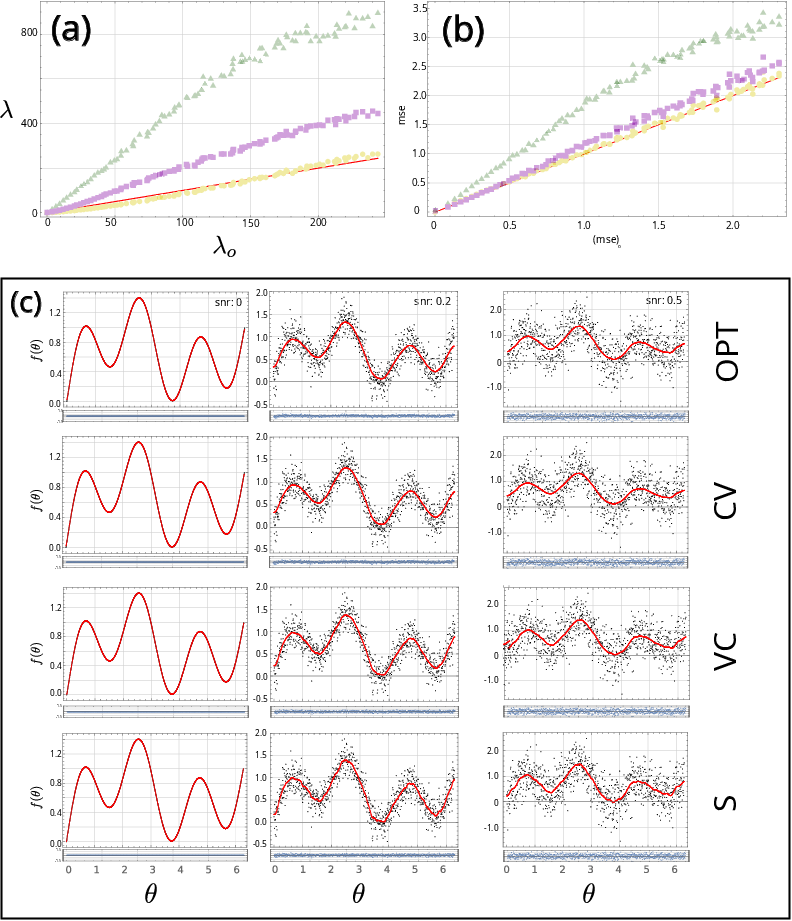}
\caption{\textbf{Noisy analytical function: $1/2(\log(t+1)+\sin t \sin 3t)$} This figure follows Figure \ref{fig:sinus}. (a) The selected $\lambda$ values obtained by the different procedures compared to the optimal $\lambda_o$. (b) The mean square error $mse$ between the non-corrupted signal $s(t)$ and the smoothed curve $\hat{s}(t)$ against the optimal mean square error. CV, VC and S correspond to the $\lambda$ value selected by the cross-validation, V-curve and spectral entropy procedure, respectively. OPT corresponds to the optimal $\lambda$ value. (c) The noisy data and the smoothed curve obtained by the different selection procedures for three noise levels (snr: 0, 0.2, and 0.5).}\label{fig:af}
\end{figure}

The same results can be observed in Figure \ref{fig:af}, which features a more complex analytical function. Again, the spectral entropy selection outperforms the cross-validation and the V-curve procedures. Cross-validation performs the worst across the entire range of noise levels. The comparison was performed several times with randomly chosen analytical functions, and in each case, the same results followed.

\subsection{Experimental data}

Finally, experimental data was used to validate the spectral entropy procedure. In this case, the mean square error can not be calculated, as the original, uncorrupted signal is unknown. Visual inspection is unavoidable for the reasons already explained in the introduction. Three datasets of different nature were used: the value of sugar stocks over time in the commodities market \cite{yahoofinance2025}; optical single-fibre spectroscopy data from the Sloan Digital Sky Survey \cite{medan25}; and nuclear magnetic resonance (NMR) data from an anonymous source. The sources of all data are public, and the data are provided in the supplementary material.

\begin{figure}[!t]
\centering
\includegraphics[scale=0.8]{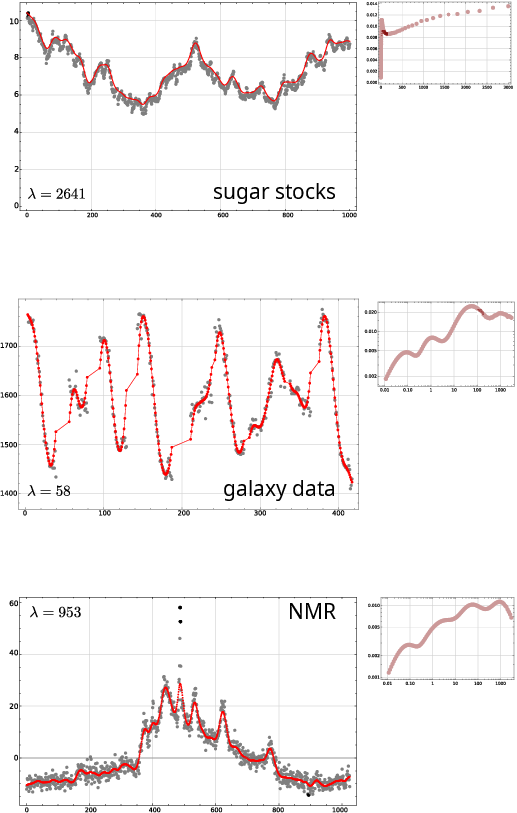}
\caption{
\textbf{Experimental Datasets.} Application and effectiveness of the proposed spectral entropy method on three distinct types of real-world experimental data. \textbf{Upper Panel (Sugar Stocks):} The data represents a time series of sugar stock prices from the commodities market. \textbf{Middle Panel (Galaxy Data):} This plot shows optical single-fiber spectroscopy data from the Sloan Digital Sky Survey. \textbf{Lower Panel (NMR Data):} The data is from a Nuclear Magnetic Resonance (NMR) experiment. Although the Eilers-Whittaker smoother is known to sometimes underestimate the intensity of very narrow peaks, the $\lambda$ selected by the spectral entropy method allows the smoothed curve to correctly identify the position of each peak's maximum. In each panel, the right inset is the S-curve obtained for its corresponsind dataset. The examples serve to validate the proposed method, showing that it produces well-behaved and visually compatible curves for data from different domains.}\label{fig:experimental}
\end{figure}

Figure \ref{fig:experimental} shows the smoothing result in each case, together with the S-curve. In all three examples, the smoothed curve resulting from the spectral entropy analysis selection yields a well-behaved curve that is compatible with the underlying noisy function. It is worth noting that the S-curve can exhibit different behaviour, and in each case, it was effective for the $\lambda$ selection.

In the case of the sugar stocks, the smoothed curve follows the overall trends in the original data, including the peaks that are clearly not a result of the noise. The galaxy data is interesting as gaps can be identified in the original experimental points. The smoothing consistently fills those gaps while following the different local trends in the experimental data. This, of course, pertains to the robust nature of the Wittaker-Eilers smoothing procedure but also suggests the adequacy of the chosen $\lambda$ value. The third data from NMR is a very noisy signal with peaks of different widths. It is known that the Wittaker-Eilers smoothing does not follow narrow peaks where more sophisticated procedures are better suited; however, even in this case, the smoothed curve correctly fits the position of each peak maxima, even if the intensity is underestimated. The selected regularization parameter yields a curve that is consistent with the underlying data, striking a good balance between smoothness and fidelity to the experimental points.

\begin{figure}[!t]
\centering
\includegraphics[scale=0.55]{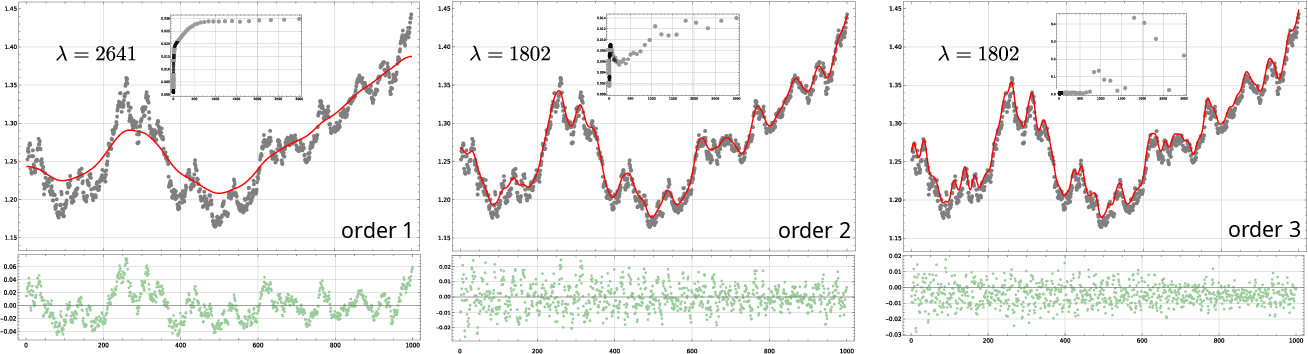}
\caption{
\textbf{Difference Operator Order.} The impact of the choice of order for the difference matrix $D$ in the behavior of the Eilers-Whittaker smoother, with the regularization parameter $\lambda$ in each case selected by the proposed spectral entropy method. The analysis is performed on a time series of Euro to Dollar conversion rates, a dataset characterized by abrupt changes and an overall increasing trend. \textbf{Order 1 (Left Panel):} The first-order difference operator produces a smoothed curve that captures the general, overall trend of the data. However, it is not sensitive enough to follow the local variations, missing the smaller peaks and troughs present in the time series. \textbf{Order 2 (Middle Panel):} Using a second-order difference operator results in a significant improvement. The smoothed curve is now able to follow all the local maxima and minima, effectively capturing the detailed fluctuations in the data while still providing a smooth and visually pleasing result. \textbf{Order 3 (Right Panel):} The third-order smoother behaves similarly to the second-order one. While it follows the local variations slightly more closely, the overall change from order 2 is not substantial. It continues to strike a good balance, capturing important details without excessively overfitting the noise in the underlying data.} \label{fig:order}
\end{figure}

The result of the smoothing procedure also depends on the order of the difference matrix. In Figure \ref{fig:order}, the smoothing result is shown for orders one, two and three. The data is a time series of Euro to Dollar conversion rates, which shows abrupt changes and a biased trend to increasing values. The order one curve gives the overall trend of the data but misses the local variations of the time series; this is fixed for the order two smoothing that can follow all local maximumns and minimums and still give a smooth curve; order three smoothing does not change much compared to the previous order, but it follows closer the local variations without overfitting to much the underlying data.

\section{Conclusions}

This paper introduced a novel procedure for selecting the optimal regularization parameter, $\lambda$, for the Eilers-Whittaker smoother, based on the spectral entropy of the residuals. This method was compared with standard techniques, such as leave-one-out cross-validation and the V-curve method, using both simulated and real-world data.

In simulations with analytical functions corrupted by Gaussian noise, the spectral entropy selection criterion consistently outperformed the other methods. The $\lambda$ values it selected were closer to the optimal $\lambda$ (the one minimizing mean square error), even at high noise levels where cross-validation and the V-curve tend to overestimate the parameter. Consequently, our method produced a smoothed signal with a mean square error consistently near the optimal minimum.

Validation with diverse experimental data—including financial, astronomical, and chemical datasets—confirmed the method's robustness. The chosen $\lambda$ parameter produced a well-behaved smoothed curve that effectively balanced noise reduction with feature preservation, handling trends, peaks, and even data gaps.

In summary, the spectral entropy selection criterion is a simple, robust, and effective tool for the automatic, unsupervised selection of the smoothing parameter. It can be a valuable addition to the existing toolkit of selection methods for the Eilers smoother.

%**********************************************************************
\section{Acknowledgments}
%**********************************************************************
%CITMA partially funded this work under the project CARDENT (PN223LH010-053).

%\bibliographystyle{unsrt}
%\bibliography{smooth.bib}

\end{document}